\documentclass[12pt]{iopart}
\usepackage[T1]{fontenc}
\usepackage[latin1]{inputenc}
\usepackage{graphicx}
\usepackage{iopams}
\usepackage{setstack}
\newcommand{\be}{\begin{equation}}
\newcommand{\ee}{\end{equation}}
\newcommand{\bea}{\begin{eqnarray}}
\newcommand{\eea}{\end{eqnarray}}
\newcommand{\bml}{\numparts}
\newcommand{\eml}{\endnumparts}

\newcommand{\vk}{\vec{k}}

\newcommand{\ep}{\epsilon}
\newcommand{\ve}{\varepsilon}

\newcommand{\op}{\mathcal{O}}
\newcommand{\opt}{\tilde{\mathcal{O}}}

\newcommand{\A}{\frac{1}{-k^2}}
\newcommand{\B}[1]{\frac{(\lambda-1){#1}}{-k^2+2(\lambda-1)\vk^2}}
\newcommand{\C}[1]{\frac{6-4\lambda}{7(3\lambda-1)}\frac{#1}{-k^2+7\frac{\lambda-1}{3\lambda-1}\vk^2}}

\begin{document}
\title{Propagator in the Ho\v rava-Lifshitz gravity}
\author{F S Bemfica and M Gomes}
\address{Instituto de Física, Universidade de São Paulo.
Caixa Postal 66318, 05315-970, São Paulo, SP, Brazil}
\eads{\mailto{fbemfica@fma.if.usp.br}, \mailto{mgomes@fma.if.usp.br}}

\begin{abstract}
In this paper it is  studied the propagator for the modified theory of gravity proposed by Ho\v rava. We first calculate the propagator in the $\lambda=1$ case and show that the main poles that arise correspond to the spin two particle and scalar particle, already known in the literature. The presence of a bad uiltraviolet behaving term spoils renormalizability of the theory but is eliminated by imposing the detailed balance condition, although just a soft version of this condition is actually needed. The problem of wrong mass sign and statistics is verified at the tree level due to the presence of the cosmological constant, demanding a complete elimination of the tadpole in order to be fully analyzed. However, in the absence of such constant the extra scalar degree of freedom has no dynamics, at least at the tree-level, and the theory posses only two dynamical degrees of freedom. Secondly, to understand the implications of $\lambda$, we analise a simplified model, the $\lambda R$ theory, and verify that the theory becomes non unitary, being a strong argument to set $\lambda=1$.
\end{abstract}

\maketitle

\section{Introduction}

Modified theories of gravity involving higher order derivative terms is an old subject. Just after the development of Einstein's general relativity (GR), proposals of its modification through the combinations of products of the Ricci tensor have been put forward~\cite{Weyl1,Eddington}. Such modifications gained strength in the context of quantum gravity, where the linearized version of Einstein's theory is known to be nonrenormalizable by power counting~\cite{Weinberg2}. Although the addition of higher order derivative terms in the action proved to be in advantage compared with pure GR due to its renormalizability character~\cite{Stelle}, it failed to be a unitarity model.

More recently, a theory of gravity with only higher order spatial derivative terms has been introduced by Ho\v rava~\cite{Horava1,Horava2}. The main idea consisted in to construct a theory with only extra higher order spatial derivative terms in the action with the aim of improving its ultraviolet behavior, with the advantage that the absence of extra time derivative should guarantee its unitarity. The foremost argument for such proposal lied in the fact that the gravity propagator of the linearized theory would behave like
\be
\label{1}
\frac{1}{\omega^2-\vk^2-a_2(\vk^2)^2-\cdots-a_z(\vk^2)^z}\,,
\ee
$a_{2},\,\cdots,\,a_z$ being coupling constants, $k^\mu=(\omega,\vk)$ the four-momentum of the graviton and $z>1$ a parameter associated with the highest order of spatial derivatives. The absence of higher order time derivatives would bring about only simple poles in $\omega^2$, as schematically written in (\ref{1}).

In a previous work~\cite{Bemfica6}, we succeeded to obtain the exact form of (\ref{1}) for a simplified Ho\v rava-Lifshitz theory with higher spatial derivatives up to the fourth order. In that case, the theory with only higher spatial derivative terms showed to be unitary, at least at the tree-level. Furthermore, no extra degrees of freedom  appeared in the propagator, whose only nonzero residue was the one corresponding to a particle with two degrees of freedom obeying a nonrelativistic dispersion relation. For this simplified model, the appearance of a bad ultraviolet term in the propagator (not a pole) has been eliminated by a detailed balance condition as the one proposed in~\cite{Horava2}, firstly introduced by Ho\v rava with the aim of avoiding the spread of the constants labeling the new extra terms. The elimination of this bad ultraviolet term showed to be in accordance with the renormalizability conditions studied in~\cite{Orlando} for the full Ho\v rava-Lifshitz theory obeying detailed balance condition.

The present work is dedicated to obtain the propagator of the full Ho\v rava-Lifshitz theory~\footnote{A study of the graviton exitation modes of different formulations of Ho\v rava gravity can be found in \cite{Helayel}.} firstly proposed in~\cite{Horava2}. The theory with up to six extra spatial derivative terms is described by the action
\bea
\label{2}
S&=&\frac{1}{\kappa^2}\int_{\Re}dt\int_{\Sigma}d^3x N\sqrt{q}\left(K^{ij}K_{ij}-\lambda K^2
+\gamma R+\alpha R^2+\beta R^{ij}R_{ij}\right.\nonumber\\
&&\left.+\,\sigma \ve^{ijk}R_{il}\nabla_j R^l_k+\delta C^{ij}C_{ij}-2\Lambda\right)\,,
\eea
defined in the foliation $\mathcal{M}\cong \Re\times\Sigma$, with index $i,\,j=1,2,3$ on $\Sigma$. $\kappa^2=16\pi G$~\footnote{In the original theory, Ho\v rava uses $\kappa^2=32\pi G$, what explains an overall fator of two}, $G$ is the Newton's constant, and, all over the paper, $c=\hbar=1$. $\lambda$ is a running parameter that must take its relativistic value $1$ in the infrared limit (also in this limit $\gamma\to1$). The totaly antisymmetric tensor $\ve^{ijk}=\ep^{ijk}/\sqrt{q}$ is written in terms of the Levi-Civita tensor density $\ep^{ijk}$, where $q=\det(q_{ij})$. The extrinsic curvature
\be
\label{3}
K_{ij}=\frac{1}{2N}\left(\dot{q}_{ij}-2\nabla_{(i}N_{j)}\right)\,,
\ee
$A_{(i}B_{j)}\equiv(A_iB_j+A_jB_i)/2$, and the Cotton tensor
\be
\label{4}
C^{ij}=\ve^{ikl}D_k\left(R_l^{(3)j}-\frac{1}{4}\delta^j_lR^{(3)}\right)
\ee
are defined by the 3-metric $q_{ij}$ on $\Sigma$, the covariant derivative $\nabla_i$ compatible with $q$, the lapse function $N$ and the shift vector $N^i$. The 3-curvature on $\Sigma$ is chosen to be $R^l_{ijk}=\partial_j\Gamma^l_{ik}-\partial_i\Gamma^l_{jk}+\cdots$ while $R_{ij}=R^l_{ilj}$. We may leave the constants $\alpha$, $\beta$, $\gamma$, $\delta$, $\sigma$, and $\Lambda$ free. However, in the original proposal~\cite{Horava2}, they take the values
\numparts
\bea
\delta&=&-\frac{\kappa^4}{\omega^4}\,,\label{5a}\\
\sigma&=&\frac{\kappa^4\mu}{\omega^2}\,,\label{5b}\\
\beta&=&-\frac{\kappa^4\mu^2}{4}\,,\label{5c}\\
\alpha&=&-\frac{4\lambda-1}{4(3\lambda-1)}\beta\,,\label{5d}\\
\gamma&=&-\frac{\kappa^4\mu^2}{4(3\lambda-1)}\Lambda_W\,,\label{5e}\\
\Lambda&=&\frac{3}{2}\gamma\Lambda_W\,,\label{5f}
\eea
\endnumparts
imposed by the detailed balance condition extracted from the 3-action
\be
\label{6}
\fl W=\mu\int_{\Sigma}d^3x\sqrt{q}\left(R^{(3)}-2\Lambda_W\right)+\frac{1}{\omega^2}\int_{\Sigma}d^3x\ve^{ijk}\left(\Gamma_{im}^l\partial_j\Gamma_{kl}^m
+\frac{2}{3}\Gamma_{il}^n\Gamma_{jm}^l\Gamma_{kn}^m\right)\,.
\ee

The present paper is organized as follows: next section will be dedicated to develop the necessary tools to calculate the propagator of the linearized version of (\ref{2}). Due to the difficulty, we first calculate the propagator for the Ho\v rava-Lifshitz theory in the $\lambda=1$ regime. We show that a partially detailed balance condition must be applied in order to eliminate a bad ultraviolet behaving term, as the one obtained in~\cite{Bemfica6}. Sec.~\ref{sec3} deals with the analises of the unitarity of the theory at the tree-level for $\lambda=1$. Although we preliminarily find poles with wrong sign in mass and a scalar particle with wrong statistics (similar results obtained in~\cite{Bogdanos}), the full treatment of the problem certainly requires a consistent elimination of the tadpole introduced by the presence of the cosmological term, also a difficulty present in pure GR~\cite{Veltman}, and is out of the scope of this paper. In Sec.~\ref{sec4} we study the modifications introduced by the $\lambda$ parameter. We use a model that mimics Ho\v rava's theory, i.e. the $\lambda R$ model~\cite{Henneaux}, and show that the theory looses its unitarity when $\lambda\ne1$. Summary and conclusions are in Sec.~\ref{summary}.

\section{The propagator}
\label{sec2}

In order to obtain the propagator in the linearized version of (\ref{2}) we may first rewrite the lapse function as $N\to N/\sqrt{\gamma}$ and define $\kappa^{\prime \,2}\equiv\kappa^2/\sqrt{\gamma}$ and also $\mho^\prime\equiv\mho/\gamma$ for $\mho=\sigma,\,\delta,\,\beta,\,\alpha,\,\Lambda$, so that the action may be cast as
\bea
\label{2.1}
\fl S=\frac{1}{\kappa^{\prime\,2}}\int_{\Re}dt\int_{\Sigma}d^3x N\sqrt{q}\left(K^{ij}K_{ij}-\lambda K^2
+ R+\alpha^\prime R^2+\beta^\prime R^{ij}R_{ij}\right.\nonumber\\
\left.+\,\sigma^\prime \ve^{ijk}R_{il}\nabla_j R^l_k+\delta^\prime C^{ij}C_{ij}-2\Lambda^\prime\right)\,.
\eea
Let us define
\numparts
\bea
g_{00}&=&-N^2+N^iN_i\,,\label{2.2a}\\
g_{0i}&=&N_i\,,\label{2.2b}\\
g_{ij}&=&q_{ij}\,.\label{2.2c}
\eea
\endnumparts
It is clear that the matrix $g_{\mu\nu}$ will become, at least in the relativistic limit where $\gamma\to1$, the metric in the manifold $\mathcal{M}$ with signature $-+++$. The definitions above enable us to write $N\sqrt{q}=\sqrt{-g}$. As claimed in~\cite{Horava2}, the action (\ref{2.1}) is still invariant under the coordinate transformations $\delta x^\mu=(\ep(t),\ep^i(t,x))$. In order to fix this gauge freedom, we choose the de Donder gauge $\Gamma^{(4)\mu}=g^{\alpha\beta}\Gamma^{(4)\mu}_{\alpha\beta}=0$ ($\Gamma^{(4)}$ is the Christopher symbol defined in terms of $g_{\mu\nu}$) by adding to the action (\ref{2.1}) the gauge fixing term
\be
\label{2.3}
S_{gf}=-\frac{\xi}{2\kappa^{\prime\,2}}\int_{\mathcal{M}}d^4x \sqrt{-g}\Gamma^{(4)\mu}\Gamma^{(4)}_\mu\,.
\ee
In the weak field approximation
\be
g_{\mu\nu}\approx \eta_{\mu\nu}+\kappa^\prime h_{\mu\nu}\,,
\ee
where $\eta_{\mu\nu}=diag(-1,1,1,1)$ is the Minkowski metric, and $\mu,\nu=0,1,2,3$, the action in (\ref{2.1}), together with the gauge fixing (\ref{2.3}), may be written, up to second order in $h$, as
\be
\label{2.4}
S_\xi\approx S^{(0)}+S^{(1)}+S^{(2)}\,.
\ee
In (\ref{2.4}), $S^{(0)}$ is an divergent constant when $\Lambda^\prime\ne0$ that may be absorbed through a redefinition of the action $S$, while
\be
\label{2.5}
S^{(1)}=-\frac{\Lambda^\prime}{\kappa^\prime}\int d^4x h
\ee
turns out to be a tadpole~\cite{Veltman}. The trace of the perturbed field has been defined as $h\equiv\eta^{\mu\nu}h_{\mu\nu}$. The conventional treatment to eliminate tadpoles shall be done by replacing
\be
\label{2.6}
h_{\mu\nu}\to h_{\mu\nu}+a_{\mu\nu}\,,
\ee
and demanding that $a_{\mu\nu}$ obeys an equation so that all terms linear in $h$ cancel. Though, the treatment of such problem turns out to be really complicated in gravity due to the fact that, contrary to what is usual in field theory, $a_{\mu\nu}$ must be a function of the coordinates~\cite{Veltman}. To illustrate this need, suppose $a_{\mu\nu}$ is a constant. Then, the only place it will show up (neglecting surface terms) in the linearized action with up to second order in $h$ will be in the square root $\sqrt{-g}$ in (\ref{2.5}) as
\bea
\label{2.7}
\fl\sqrt{-\det(\eta+h+a)}\approx\sqrt{-\det(\eta+a)}\left[1+\frac{1}{2}D^{\mu\nu}h_{\mu\nu}\right.\nonumber\\
\fl\phantom{\sqrt{-\det(\eta+h+a)}\approx}\left.-\frac{1}{4}D^{\mu\alpha}h^\nu_\alpha D_{\mu\beta}
h^\beta_\nu+\frac{1}{8}\left(D^{\mu\nu}h_{\mu\nu}\right)^2\right]\,,
\eea
where the matrix $D=(I+\eta a)^{-1}$ ($I$ is the $4\times4$ identity matrix and, in that equation only, $\eta$, $h$, and $a$ should be understood as matrixes, not traces). Cancellation of the linear terms in $h_{\mu\nu}$ demands that $\det(D)=0$ [$\det(I+\eta a)\to\infty$], what is inconsistent. A complete treatment of the tadpole is too complicated and needs all orders in $a$ even though we are limiting ourselves to order two in $h$. We will return to this point when discussing the propagator and the problem of unitarity.

Let us now concentrate in the term $S^{(2)}$. It can be written in the quadratic form
\be
\label{2.8}
S^{(2)}=\frac{1}{2}\int d^4x h^{\mu\nu}\op_{\mu\nu,\alpha\beta}h^{\alpha\beta}\,.
\ee
Clearly, the operator $\op$ possesses the symmetries $\op_{\mu\nu,\alpha\beta}=\op_{\alpha\beta,\mu\nu}=\op_{\nu\mu,\alpha\beta}$. For convenience, as done in~\cite{Bemfica6}, we split
\be
\label{2.9}
\op_{\mu\nu,\alpha\beta}=A_{\mu\nu,\alpha\beta}+\delta_{\mu\nu}^{ij}B_{ij,kl}\delta_{\alpha\beta}^{kl}\,,
\ee
where $\delta^{\mu\nu}_{\alpha\beta}\equiv \delta^\mu_{(\alpha}\delta^\nu_{\beta)}$, $A_{\mu\nu,\alpha\beta}$ does not contain pure spatial indices, while $B_{ij,kl}$ is the spatial sector of $\op$. In momentum space $k^{\mu\nu}=(\omega,\vk)$,
\numparts
\bea
A_{00,00}&=&-\frac{\xi}{4}k^2+\frac{\Lambda^\prime}{2}\,,\label{2.10a}\\
A_{00,ij}&=&\frac{\delta_{ij}}{2}\left[\frac{\xi}{2}k^2-k^2-(1-\xi)\omega^2+\Lambda^\prime\right]+\frac{1-\xi}{2}k_ik_j\,,\label{2.10b}\\
A_{ij,0k}&=&\frac{\omega}{2}(1-\xi)\left(-\delta_{k(i}k_{j)}+k_k\delta_{ij}\right)+\frac{\lambda-1}{2}\omega\delta_{ij}k_k\,,\label{2.10c}\\
A_{0i,0j}&=&\frac{\delta_{ij}}{4}\left[k^2+(1-\xi)\omega^2-2\Lambda^\prime\right]-\frac{2\lambda-1-\xi}{4}k_ik_j\equiv M_{ij}\,,\label{2.10d}
\eea
\endnumparts
and zero otherwise. It must be clear from the symmetries of $\opt$ that $A_{\mu\nu,\alpha\beta}=A_{\alpha\beta,\mu\nu}=A_{\nu\mu,\alpha\beta}$. In the spatial sector
\bea
\label{2.11}
\fl B_{ij,kl}=\delta_{ij,kl}\left(-\frac{k^2}{2}+\frac{\beta^\prime}{2}\vk^4-\frac{\delta^\prime}{2}\vk^6+\Lambda^\prime\right)\nonumber\\
+\delta_{ij}\delta_{kl}\left[-\frac{\lambda}{2}\omega^2-\frac{\xi}{4} k^2 +\frac{\vk^2}{2}+\left(2\alpha^\prime+\frac{\beta^\prime}{2}\right)\vk^4 + \frac{\delta^\prime}{4}\vk^6-\frac{\Lambda^\prime}{2}\right]\nonumber\\
+\delta_{((i(k}k_{l)}k_{j))}\left(-\xi+1-\beta^\prime\vk^2+\delta^\prime\vk^4\right)\nonumber\\
+\left(\delta_{ij}k_kk_l+\delta_{kl}k_ik_j\right)\left[\frac{\xi-1}{2}-\frac{\delta^\prime}{4}\vk^4-
\left(2\alpha^\prime+\frac{\beta^\prime}{2}\right)\vk^2\right]\nonumber\\
+k_ik_jk_kk_l\left[\left(2\alpha^\prime+\frac{\beta^\prime}{2}\right)-\frac{\delta^\prime}{4}\vk^2\right]
+2i\sigma^\prime \mathbb{P}_{ij,kl}\,,
\eea
where
\be
\label{2.12}
\mathbb{P}_{ij,kl}\equiv \frac{\vk^4}{4}\ep_{((im(k}k^m\theta_{l)j))}\,,
\ee
and is defined in terms of the tree-dimensional transverse projector $\theta_{ij}$ defined in (\ref{24}). The double bracket   $A_{((i(k}B_{l)j))}\equiv(A_{i(k}B_{l)j}+A_{j(k}B_{l)i})/2$ is used in the same sense as the single one defined somewhere in the text.

To obtain the propagator $\opt^{-1}$ in momentum space we must solve the equation
\be
\label{2.13}
\opt_{\mu\nu,\alpha\beta}\opt^{-1\,\alpha\beta,\lambda\delta}=\delta^{\lambda\delta}_{\mu\nu}\,.
\ee
By splitting the above equation into pure spatial indexes and mixed indexes, the relevant equations to solve turn out to be
\numparts
\bea
C_{ij,kl}\opt^{-1\,kl,mn}=\delta^{mn}_{ij}\,.\label{2.14a}\\
\opt^{-1\,0i,mn}=-\frac{1}{2}M^{-1\,ij}A_{0j,kl}\opt^{-1\,kl,mn}\,,\label{2.14b}\\
\opt^{-1\,00,mn}=-\frac{A_{00,kl}}{A_{00,00}}\opt^{-1\,kl,mn}\,,\label{2.14c}\\
\opt^{-1\,0i,00}=-\frac{1}{2}M^{-1\,iq}A_{0q,kl}\opt^{-1\,kl,00}\,,\label{2.14d}\\
\opt^{-1\,0i,0m}=\frac{1}{4}M^{-1\,im}-\frac{1}{2}M^{-1\,ij}A_{0j,kl}\opt^{-1\,kl,0m}\,,\label{2.14e}\\
\opt^{-1\,00,00}=\frac{1}{A_{00,00}}\left(1-A_{00,ij}\opt^{-1\,ij,00}\right)\,,\label{2.14f}
\eea
\endnumparts
where, from (\ref{2.10d}),
\be
\label{2.15}
M^{-1\,ij}=\frac{4}{k^2+(1-\xi)\omega^2}\left(\delta^{ij}+\frac{(2\lambda-1-\xi)}{\xi k^2+2(1-\lambda)\vk^2}\right)\,.
\ee
To obtain $C_{ij,kl}$ in (\ref{2.14a}), one must fix indexes in (\ref{2.13}) as follows
\bml
\bea
\opt_{ij,\mu\nu}\opt^{-1\,\mu\nu,kl}=2A_{ij,0\mu}\opt^{-1\,0\mu,kl}+B_{ij,mn}\opt^{-1\,mn,kl}=\delta^{kl}_{ij}\,.
\eea
\eml
The $\opt^{-1\,0\mu,kl}$ terms were obtained also by fixing free indices in (\ref{2.13}) and are written in (\ref{2.14b}) and (\ref{2.14c}). Collecting those results one gets
\bea
\label{2.17}
\fl C_{ij,kl}&=&-\frac{A_{00,kl}A_{ij,00}}{A_{00,00}}-A_{ij,0m}M^{-1\,mn}A_{0n,kl}+B_{ij,kl}\nonumber\\
\fl&=&\delta_{ij,kl}\frac{1}{2}\left(-k^2+\beta^\prime\vk^4-\delta^\prime\vk^6+2\Lambda^\prime\right)\nonumber\\
\fl&+&\delta_{ij}\delta_{kl}
\left[\frac{1}{2}\left(\vk^2-\lambda\omega^2\right)-\frac{\xi}{4}k^2
+\left(2\alpha^\prime+\frac{\beta^\prime}{2}\right)\vk^4+\frac{\delta^\prime}{4}\vk^6-\frac{I^2}{H}-\frac{L^2\vk^2}{F+G\vk^2}\right]\nonumber\\
\fl&+&\frac{4}{\vk^2}\delta_{((i(k}k_{l)}k_{j))}\left[(1-\xi-\beta^\prime\vk^2+\delta^\prime\vk^4)\frac{\vk^2}{4}-\frac{J^2\omega^2\vk^2}{4F}\right]\nonumber\\
\fl&+&\frac{\delta_{ij}k_kk_l+\delta_{kl}k_ik_j}{\vk^2}\left[\frac{1}{2}(\xi-1)\vk^2
-\left(2\alpha^\prime+\frac{\beta^\prime}{2}\right)\vk^4-\frac{IJ\vk^2}{H}+\frac{JL\omega\vk^2}{F+G\vk^2}-\frac{\delta^\prime}{4}\vk^6\right]\nonumber\\
\fl&+&\frac{k_ik_jk_lk_k}{\vk^4}\left[\left(2\alpha^\prime+\beta^\prime-\frac{J^2}{H}+\frac{GJ^2\omega^2}{F\left(F+G\vk^2\right)}\right)\vk^4
-\frac{\delta^\prime}{4}\vk^6\right]+2i\sigma^\prime \mathbb{P}_{ij,kl}\,.\nonumber\\
\fl
\eea
To shorten the above equation we defined
\bml
\bea
F&=&\frac{k^2+(1-\xi)\omega^2}{4}-\frac{\Lambda^\prime}{2}\,,\label{2.18a}\\
G&=&-\frac{2\lambda-1-\xi}{4}\,,\label{2.18b}\\
H&=&-\frac{\xi}{4}k^2+\frac{\Lambda^\prime}{2}\,,\label{2.18c}\\
I&=&\frac{\xi-2}{4}k^2+\frac{\xi-1}{2}\omega^2+\frac{\Lambda^\prime}{2}\,,\label{2.18d}\\
J&=&\frac{1-\xi}{2}\,,\label{2.18e}\\
L&=&\frac{\lambda-\xi}{2}\omega\,.\label{2.18f}
\eea
\eml

The tools to calculate the spatial sector $\opt^{-1}_{ij,kl}$ is developed in \ref{ApendiceA}. However, in the case $\lambda\ne1$, it turns out to far complicated. To understand what is going on, first we will limit ourselves to the case $\lambda=1$. Latter we will come back to the general case $\lambda\ne1$ for the simplified model worked in~\cite{Henneaux}.

\subsection{The special case $\lambda=1$}

We now turn to the problem of finding the complete propagator $\opt^{-1}$ in momentum space for the case $\lambda=1$. We also may choose the gauge $\xi=1$, for simplicity. Then, we are ready to obtain the spatial index sector of the propagator that may be acquired by rewriting $C$ given in (\ref{2.17}) in the form of that in (\ref{25-1}) by applying Eqs.~(\ref{25a})--(\ref{25e}). After that, and by making use of the inverse formula (\ref{25-2}), one finds that
\bea
\label{2.19}
\fl \opt^{-1}_{ij,kl}&=&\frac{\left(2P^1+\bar{P}^0-\bar{\bar{P}}^0\right)_{ij,kl}}{-k^2+2\Lambda^\prime}
+y_2P^2_{ij,kl}+y_3\mathbb{P}_{ij,kl}-\left(8\alpha^\prime+3\beta^\prime\right)\frac{\vk^4}{(k^2-2\Lambda^\prime)}\bar{P}^0\nonumber\\
\fl &=&\frac{\left(2\mathcal{P}^1+2\mathcal{P}^2-\mathcal{P}^0+\bar{\mathcal{P}}^0-\bar{\bar{\mathcal{P}}}^0\right)_{ij,kl}}{-k^2+2\Lambda^\prime}
+P^2_{ij,kl}\left(y_2-\frac{2}{-k^2+2\Lambda^\prime}\right)\nonumber\\
\fl &&+y_3\mathbb{P}_{ij,kl}-\left(8\alpha^\prime+3\beta^\prime\right)\frac{\vk^4}{(-k^2+2\Lambda^\prime)^2}\bar{P}^0_{ij,kl}\,.
\eea
From the first to the second equality in the above equation we recurred to the identities (\ref{28a}) and (\ref{28b}). Also, we defined
\bml
\bea
y_2&\equiv&\frac{1}{-k^2+\beta^\prime\vk^4+\sqrt{2}|\sigma^\prime\vk^5|-\delta^\prime\vk^6+2\Lambda^\prime }\nonumber\\
&&+\frac{1}{-k^2+\beta^\prime\vk^4-\sqrt{2}|\sigma^\prime\vk^5|-\delta^\prime\vk^6+2\Lambda^\prime }\,,\label{2.20a}\\
y_3&\equiv&-\frac{4i\sigma^\prime}{2\left(-k^2+\beta^\prime\vk^4-\delta^\prime\vk^6+2\Lambda^\prime\right)^2-\sigma^{\prime\,2}\vk^{10} }\,.\label{2.20b}
\eea
\eml
The remaining terms of the total propagator are obtained by substituting (\ref{2.19}) into equations (\ref{2.14b})--(\ref{2.14f}), also invoking the properties of $\mathbb{P}_{ij,kl}$ worked in \ref{ApendiceA}. The full propagator is, then,
\bea
\label{2.21}
\fl \opt^{-1}_{\mu\nu,\alpha\beta}&=&\frac{\left(2\mathcal{P}^1+2\mathcal{P}^2-\mathcal{P}^0
+\bar{\mathcal{P}}^0-\bar{\bar{\mathcal{P}}}^0\right)_{\mu\nu,\alpha\beta}
-2\delta_{\mu\nu}^{ij}\delta_{\alpha\beta}^{kl}\,P^{2}_{ij,kl}}{-k^2+2\Lambda^\prime}\nonumber\\
\fl&&-\frac{\left(8\alpha^\prime+3\beta^\prime\right)\vk^4\mathcal{Q}_{\mu\nu,\alpha\beta}}{\left(-k^2+2\Lambda^\prime\right)^2}+\delta_{\mu\nu}^{ij}\delta_{\alpha\beta}^{kl}\,P^{2}_{ij,kl}
\left(\frac{1}{-k^2+\beta^\prime\vk^4+\sqrt{2}|\sigma^\prime\vk^5|-\delta^\prime\vk^6+2\Lambda^\prime }\right.\nonumber\\
\fl&&\left.+\frac{1}{-k^2+\beta^\prime\vk^4-\sqrt{2}|\sigma^\prime\vk^5|-\delta^\prime\vk^6+2\Lambda^\prime }
\right)\nonumber\\
\fl&&-\frac{4i\sigma^\prime\delta_{\mu\nu}^{ij}\delta_{\alpha\beta}^{kl}\mathbb{P}_{ij,kl}}{2\left(-k^2+\beta^\prime\vk^4-\delta^\prime\vk^6+2\Lambda^\prime\right)^2-\sigma^{\prime\,2}\vk^{10}}\,,
\eea
where,
\bea
\label{2.22}
\mathcal{Q}_{\mu\nu,\alpha\beta}\Longrightarrow
\cases{\mathcal{Q}_{ij,kl}=\bar{P}^0_{ij,kl}\,,\\
\mathcal{Q}_{00,00}=-1\,,\\
\mathcal{Q}_{00,mn}=\frac{k_mk_n}{\vk^2}=\mathcal{Q}_{mn,00}\,,\\
0,\quad\mathrm{otherwise}.}
\eea

Notice that the relevant poles
\bml
\bea
\omega^2&=&\vk^2-2\Lambda^\prime\,,\label{2.23a}\\
\omega^2_{\pm}&=&\vk^2-\beta^\prime\vk^4\pm\sqrt{2}|\sigma^\prime\vk^5|+\delta^\prime\vk^6-2\Lambda^\prime\label{2.23b}
\eea
\eml
have the wrong sign in the mass if $\Lambda^\prime>0$. It is not the case if we impose detailed balance condition, where $\Lambda_W<0$ because $\gamma>0$, since it is related to the emergent light speed~\cite{Horava2}. Though, we have not treated the tadpole (\ref{2.5}) yet. We know from field theory that the elimination of tadpoles may change masses so that we hope this problem should be resolved after that. This is also a problem in pure GR with cosmological constant~\cite{Veltman}. The wrong mass and statistic problem in the full Ho\v rava-Lifshitz theory has been raised in~\cite{Bogdanos}, although its correction through the tadpole elimination has not been considered.

In the propagator obtained above, we follow the interpretation given in~\cite{Bogdanos} where the two distinct poles in (\ref{2.23b}) are said to correspond to the same spin two particle ($P^2$ has only two independent indexes)  with two different polarizations. The reader may also notice that the pure GR pole (\ref{2.23a}) with cosmological constant $\Lambda^\prime$ has been reobtained in the gauge $\xi=1$, contrary to the expected form (\ref{1}) to the propagator. However, it has suffered a correction proportional to $P^2$ whose consequences shall be understood in the next section, together with the other terms in the propagator (\ref{2.21}). In ref.~\cite{Bemfica6}, in the absence of a cosmological constant, we showed that such pole does not have dynamics, at least at the tree level.

It is worth mentioning that the term containing $\mathcal{Q}$ has a bad ultraviolet behavior, since by power counting in momentum $\vk$ it has a worse behavior than that of pure gravity. In~\cite{Bemfica6} we have faced the same problem. However, the imposition of a detailed balance condition in that situation have canceled this undesired term. In the present case we got a similar situation, where the detailed balance condition written in (\ref{5d}), for $\lambda=1$, lead us to $8\alpha^\prime+3\beta^\prime=0$, eliminating the only bad ultraviolet term in the propagator (\ref{2.21}). It is also interesting to note that such problem raises only when in the presence of the terms $R^2$ and $R^{ij}R_{ij}$, whatever combination we choose for the remaining extra terms in (\ref{2}). This suggests that, regarding renormalizability of the theory, there is a possibility of relaxation of the detailed balance condition provided we keep $\alpha^\prime=-3\beta^\prime/8$.

\section{Tree-level unitarity}
\label{sec3}

At this point, we need to verify the physical degrees of freedom in the propagator (\ref{2.21}) and test its unitarity at the tree level. To do this, we must saturate the propagator with an arbitrary conserved current $T^{\mu\nu}$ ($k_\mu \tilde{T}^{\mu\nu}=0$ in momentum space)~\cite{Veltman,Nieuwenhuizen,Accioly}. The most general conserved current can be spanned by the four linearly independent vectors $k^\mu=(\omega,\vk)$, $\tilde{k}^\mu=(-\omega,\vk)$, and $\ep^{\mu}_r=(0,\vec{\ep}_r)$ with $\vec{\ep}_r\cdot\vec{\ep}_s=\delta_{rs}$, where $r,s=1,2$ corresponds to the two graviton transverse directions, i.e., $\vec{\ep}_r\cdot\vk=0$. Arbitrarily we write
\be
\label{3.1}
\tilde{T}^{\mu\nu}(k)=ak^\mu k^\nu+b\tilde{k}^\mu\tilde{k}^\nu+ c_{rs}\ep^\mu_r\ep^\nu_s+2d\tilde{k}^{(\mu}k^{\nu)}+2e_r\ep^{(\mu}_r k^{\nu)}
+2f_r\ep_r^{(\mu}\tilde{k}^{\nu)}\,,
\ee
with an explicit sum over repeated indexes $r$ and $s$. There are $10$ arbitrary coefficients $a,b,c_{rs},d,e_r,f_r$ ($c_{rs}=c_{sr}$) that reduces to the six independent components of the conserved current by requiring its conservation $k_\mu\tilde{T}^{\mu\nu}=0$, i.e.,
\bml
\bea
&&ak^2+d(\omega^2+\vk^2)=0\,,\label{3.11a}\\
&&dk^2+b(\omega^2+\vk^2)=0\,,\label{3.11b}\\
&&e_rk^2+f_r(\omega^2+\vk^2)=0\,.\label{3.11c}
\eea
\eml
Reality condition on $T^{\mu\nu}(x^\mu)$ requires $\tilde{T}^{\mu\nu}(-k)=\tilde{T}^{*\mu\nu}(k)$ and, as a consequence, $a,b,c_{ab},d\in\Re$, while $e_r$ and $f_r$ are pure imaginary. From now on we are going to impose $8\alpha^\prime+\beta^\prime=0$ to eliminate the bad ultraviolet term in the propagator. Let us define the amplitude
\bea
\label{3.2}
\fl\mathcal{A}=T^{*\mu\nu}(k)\opt^{-1}_{\mu\nu,\alpha\beta}T^{\alpha\beta}(k)
=\frac{k^2(b-a)\left[(b-a)k^2-2c_{rr}\right]+4f_a^*\left[f_rk^2+e_r(\omega^2+k^2)\right]}{-k^2+2\Lambda^\prime}\nonumber\\
+\left(2c_{rs}c_{rs}-c_{rr}^2\right)
\left(\frac{1}{-k^2+\beta^\prime\vk^4+\sqrt{2}|\sigma^\prime\vk^5|-\delta^\prime\vk^6+2\Lambda^\prime }\right.\nonumber\\
\left.+\frac{1}{-k^2+\beta^\prime\vk^4-\sqrt{2}|\sigma^\prime\vk^5|-\delta^\prime\vk^6+2\Lambda^\prime }
\right)\,,
\eea
where (\ref{3.11a})--(\ref{3.11c}) have been taken into account. The elimination of $\mathbb{P}$, given by
\be
\label{3.3}
T^{*ij}\mathbb{P}_{ij,kl}T^{kl}=\frac{\vk^4}{4}c_{rs}c_{st}(\vec{\ep}_r\times\vec{\ep}_t)^mk^m=0\,,
\ee
was obtained applying its properties described in \ref{ApendiceA}. The residues of the remaining poles turn out to be:
\begin{itemize}
\item Pole $-k^2+2\Lambda^\prime=0$
\be
\label{3.4}
Res\,\mathcal{A}_{\omega}=4\Lambda^\prime(b-a)\left[(b-a)\Lambda^\prime-c_{rr}\right]+8f_r^*\left[f_r\Lambda^\prime+e_r(\vk^2-\Lambda^\prime)\right]\,;
\ee
\item Poles $-k^2+\beta^\prime\vk^4\pm\sqrt{2}|\sigma^\prime\vk^5|-\delta^\prime\vk^6+2\Lambda^\prime=0$
\be
\label{3.5}
Res\,\mathcal{A}_{\omega_{\pm}}=\left(c^{11}-c^{22}\right)^2+4(c^{12})^2\ge0\,.
\ee
\end{itemize}

Notice that the residue of the pole $-k^2+2\Lambda^\prime=0$ is not strictly positive, since it depends on the value of $\Lambda^\prime$ as well as $\vk$. In the absence of the cosmological constant $\Lambda^\prime$ it clearly eliminates such problem and also switch the residue of this pole to zero, because $f_r=0$ when $k^2=0$ from (\ref{3.11c}). This is what have occurred in our previous work~\cite{Bemfica6}. This means that we can eliminate the dynamics of the scalar degree of freedom by discarding the cosmological constant. In the case of keeping the cosmological constant, one must eliminates the tadpole in (\ref{2.5}) and only then analise the residue of this problematic pole to verify the sign of its residue. The new pole corresponding to the spin two particle with the two polarizations $\omega_{\pm}$ has the right statistics. The problem of mass in this pole also probably relies in the tadpole elimination process.

\section{The $\lambda R$ model}
\label{sec4}

Last section we focused on the analises of the full Ho\v rava theory with $\lambda=1$ due to the difficulty generated by an arbitrary $\lambda$. Now we concentrate in the $\lambda\ne1$ case, again in the gauge $\xi=1$, for a simplified model in order to understand the main modifications introduced by this parameter in the propagator. The $\lambda R$ model, first studied in~\cite{Henneaux} in the context of constraint analises in the Hamiltonian formalism, is described by the action
\be
\label{4.1}
S=\frac{1}{\kappa^2}\int_\Re dt\int_\Sigma d^3x N\sqrt{q}\left(K^{ij}K_{ij}-\lambda K^2+R\right)\,,
\ee
and corresponds to the case $\gamma=1$, $\sigma=\delta=\beta=\alpha=0$ in the present work. By adjusting such parameters in the equations in Sec.~\ref{sec2} we readily obtain, for the spatial sector of the propagator,
\bea
\label{4.2}
\fl\opt^{-1}_{ij,kl}&=&\frac{\left(2P^1+2P^2+\frac{2}{7}P^0+\frac{8}{7}\bar{P}^0-\frac{6}{7}\bar{\bar{P}}^0\right)_{ij,kl}}{-k^2}
-\frac{6-4\lambda}{7(3\lambda-1)}\frac{\left(2P^0+\bar{P}^0+\bar{\bar{P}}^0\right)_{ij,kl}}{-k^2+7\frac{\lambda-1}{3\lambda-1}\vk^2}\nonumber\\
\fl&=&\frac{\left(2\mathcal{P}^1+2\mathcal{P}^2-\frac{4}{7}\mathcal{P}^0+\frac{8}{7}\bar{\mathcal{P}}^0-\frac{6}{7}\bar{\bar{\mathcal{P}}}^0\right)_{ij,kl}}{-k^2}
-\frac{6-4\lambda}{7(3\lambda-1)}\frac{\left(3\mathcal{P}^0+\bar{\mathcal{P}}^0+\bar{\bar{\mathcal{P}}}^0\right)_{ij,kl}}{-k^2+7\frac{\lambda-1}{3\lambda-1}\vk^2}\,,
\eea
where, from the first to the second equality, we used the relations (\ref{28a}) and \eref{28b}. Again, by inserting the above result into equations (\ref{2.18b})--(\ref{2.18f}) one gets
\be
\label{4.3}
\opt^{-1}_{\mu\nu,\alpha\beta}=
\frac{\left(2\mathcal{P}^1+2\mathcal{P}^2-\mathcal{P}^0+\bar{\mathcal{P}}^0-\bar{\bar{\mathcal{P}}}^0\right)_{\mu\nu,\alpha\beta}}{-k^2}
+\mathcal{U}_{\mu\nu,\alpha\beta}\,.
\ee
Once more, the theory contains the GR propagator with pole $-k^2=0$, but without cosmological constant. The presence of $\lambda$ have generated the noncovariant term
\be
\label{4.4}
\fl\mathcal{U}_{\mu\nu,\alpha\beta}\Rightarrow
\cases{
\mathcal{U}_{00,00}=\frac{9}{7}\frac{1}{-k^2}-\C{9}\\
\mathcal{U}_{0i,mn}=-\B{\omega k_i\delta_{mn}}\left(\frac{4}{7}\A+\C{3}\right)\\
\mathcal{U}_{00,mn}=3\delta_{mn}\left(-\frac{1}{7}\A+\C{1}\right)\\
\mathcal{U}_{0i,00}=\B{\omega k_i}\left(\frac{12}{7}\A+\C{9}\right)\\
\mathcal{U}_{0i,0j}=\B{k_ik_j}\left[\rule{0cm}{.7cm}\right.\frac{2}{-k^2}-\B{\omega^2}\left(\rule{0cm}{.7cm}\right.\frac{12}{7}\A\\
\phantom{\mathcal{U}_{0i,0j}=\B{k_ik_j}\left[\rule{0cm}{.7cm}\right.}\left.\left.+\C{9}\right)\right]\\
\mathcal{U}_{ij,kl}=\left(3\mathcal{P}^0+\bar{\mathcal{P}}^0+\bar{\bar{\mathcal{P}}}^0\right)_{ij,kl}\left(\frac{1}{7}\frac{1}{-k^2}
-\frac{6-4\lambda}{7(3\lambda-1)}\frac{1}{-k^2+7\frac{\lambda-1}{3\lambda-1}\vk^2}\right)}.
\ee
Apparently, there is a spread of poles, including a quadratic one $\left[-k^2+2(\lambda-1)\vk^2\right]^2$. To study the physical degrees of freedom and check unitarity at the tree-level, we again saturate the propagator with the conserved current $\tilde{T}^{\mu\nu}$ given in (\ref{3.1}). Let us again define
\be
\label{4.5}
\mathcal{A}=\tilde{T}^{*\mu\nu}\opt_{\mu\nu,\alpha\beta}\tilde{T}^{\mu\nu}=\mathcal{A}^{GR}+\tilde{\mathcal{A}}\,,
\ee
where we have separated the pure GR amplitude
\bea
\label{4.6}
\fl\mathcal{A}^{GR}&=&\frac{\tilde{T}^{*\mu\nu}\left(2\mathcal{P}^2-\mathcal{P}^0+\bar{\mathcal{P}}^0-\bar{\bar{\mathcal{P}}}^0\right)\tilde{T}^{\alpha\beta}}{-k^2}\nonumber\\
\fl&=&\frac{2c_{rs}c_{rs}-c_{rr}^2+k^2(b-a)\left[(b-a)k^2-2c_{rr}\right]+4f_a^*\left[f_rk^2+e_r(\omega^2+k^2)\right]}{-k^2}
\eea
from the modifications introduced by $\lambda$ contained in
\bea
\label{4.7}
\fl\tilde{\mathcal{A}}&=&\tilde{T}^{*\mu\nu}\mathcal{U}_{\mu\nu,\alpha\beta}\tilde{T}^{\alpha\beta}=
\frac{\omega^2(a-b)^2\left[-4k^2+\frac{8\omega^2(\lambda-1)}{(3-2\lambda)}\right]}{-k^2+2(\lambda-1)\vk^2}\nonumber\\
\fl&+&\frac{\frac{1}{7}\left[3\omega^2(a+b-2d)-(a+b+2d)\vk^2-c_{rr}+4\omega^2(a-b)\right]^2+4k^2\omega^2(a-b)^2}{-k^2}\nonumber\\
\fl&-&\frac{\frac{6-4\lambda}{7(3\lambda-1)}\left[3\omega^2(a+b-2d)-(a+b+2d)\vk^2-c_{rr}+4\omega^2(a-b)\frac{3\lambda-1}{6-4\lambda}\right]^2}{-k^2+7\frac{\lambda-1}{3\lambda-1}}\,,
\eea
obtained by carefully summing $\tilde{T}^{*\mu\nu}\mathcal{U}_{\mu\nu,\alpha\beta}\tilde{T}^{\alpha\beta}=
\tilde{T}^{*00}\mathcal{U}_{00,00}\tilde{T}^{00}+2\tilde{T}^{*00}\mathcal{U}_{00,ij}\tilde{T}^{ij}+4\tilde{T}^{*00}\mathcal{U}_{00,0i}\tilde{T}^{0i}+\cdots$ using (\ref{4.4}) together with (\ref{3.1}). The interesting result is that the quadratic pole $\left[-k^2+2(\lambda-1)\vk^2\right]^2$ has been canceled. Conditions (\ref{3.11a})--(\ref{3.11c}) enable one to find
\be
\label{4.8}
Res(\mathcal{A}^{GR})_{-k^2=0}=\left(c^{11}-c^{22}\right)^2+4(c^{12})^2\ge0\,.
\ee
The remaining residues, including the correction in the residue of the pole $-k^2=0$ are:
\begin{itemize}
\item Pole $\omega^2=\vk^2$. In this case, $b=d=f_a=0$. The result is, then,
\be
\label{4.9}
Res(\tilde{\mathcal{A}})_{-k^2=0}=\frac{1}{7}\left[6\vk^2a-c_{rr}\right]^2\ge0\,.
\ee
This modification is healthy since it cannot change the sign of the residue of this pole when combined with (\ref{4.8}).
\item Pole $\omega^2=\vk^2-2(\lambda-1)\vk^2=(3-2\lambda)\vk^2\Longrightarrow \lambda\le 3/2$. The result is
\be
\label{4.10}
Res(\tilde{\mathcal{A}})_{-k^2+2(\lambda-1)\vk^2=0}=0\,.
\ee
In other words, this pole has no dynamics, at least at the tree level. This is fine, since we do not want a spread of degrees of freedom.
\item Pole $\omega^2=\vk^2-7(\lambda-1)\vk^2/(3\lambda-1)=(6-4\lambda)\vk^2/(3\lambda-1)$. Again, reality implies $\lambda\in (1/3,3/2]$. Eqs.~(\ref{3.11a})--(\ref{3.11c}) lead us to the relations $d=-7a(\lambda-1)/(3\lambda-1)$ and $b=49a(\lambda-1)^2/(3\lambda-1)^2$, with the resulting residue
\be
\label{4.11}
\fl Res(\tilde{\mathcal{A}})_{-k^2+7\vk^2\frac{\lambda-1}{3\lambda-1}=0}=-\frac{6-4\lambda}{7(3\lambda-1)}
\left[\frac{8a\vk^2(3-2\lambda)(25\lambda-13)}{(5-\lambda)^2}-c_{rr}\right]^2\le0\,.
\ee
The extra degree of freedom in the above equation clearly corresponds to a ghost unless the two following situations apply: $\lambda=1$, which is pure GR; $\lambda=3/2$, where the corresponding particle has no dynamics. However, we expect that $\lambda$ is dynamical and must run to $1$ in the infrared limit. If $\lambda$ runs continuously, in this process this ghost will acquire dynamics and will be harmful. The theory in the presence of the parameter $\lambda$ demands extra corrections. A consistent exention to the Ho\v rava gravity in such case can be found in Ref.~\cite{Blas2}.
\end{itemize}

\section{Summary and Conclusions}
\label{summary}

In this work we have obtained and analyzed the propagator for the Ho\v rava-Lifshitz gravity. To begin with, in Sec.~\ref{sec2} we limited ourselves to the case $\lambda=1$ and showed that the corresponding propagator contains the following: a pole corresponding to the scalar excitation; a spin two particle with two distinct polarizations corresponding to two different dispersion relations, possibly with wrong sign in mass depending on the sign of the cosmological constant; a bad ultraviolet behaving term that must spoil renormalizability; and also a nonphysical pole. The bad ultraviolet behaving term is automatically eliminated by imposing the detailed balance condition or, as the case worked in~\cite{Bemfica6}, a partial detailed balance condition is enough because this problem arises only in the presence of the terms $R^2$ and $R^{ij}R_{ij}$.

In Sec.~\ref{sec3}, we studied the tree-level unitarity and, as a result, the nonphysical pole was automatically eliminated. One problem arises due to the fact that the pole corresponding to the scalar excitation has no positive defined residue. This problem was also raised in~\cite{Bogdanos}, together with the wrong mass sign problem. However, none of us have treated the tadpole in the theory. We argue that a careful elimination of such linear term in the field $h$ could change masses and also cure the problem of negative residues. This is not a problem restrict to the Ho\v rava gravity and is also present in pure general relativity with cosmological constant~\cite{Veltman}. One can see that the elimination of the cosmological constant eliminates this problem, and also eliminates the dynamics of this scalar particle, reducing the number of degrees of freedom to the desired value of two. However, a complete treatment of the tadpole seems to be extremely complicated and we did no addressed to this problem.

Section~\ref{sec4} was dedicated to study the general case $\lambda\ne1$. Due to the difficulties of treating the full Ho\v rava-Lifshitz theory in the presence of $\lambda$, we limited ourselves to the simplified $\lambda R$ theory studied first in~\cite{Henneaux} in the context of constraint analises. After we calculate the propagator of this model, we showed that, beyond the limiting range $\lambda\in (1/3,3/2]$ for this parameter, the theory turns out to be non-unitary. The only values of $\lambda$ for which the theory ir unitary is $1$ and $3/2$. Nevertheless, if $\lambda$ runs continuously to $1$ in the infrared limit, the theory is not unitary anyway. The difference between the $\lambda R$ and Ho\v rava-Lifshitz complete theory relies in the extra spatial derivative term and the ghost present in the first will probably be present in the second.

\ack
The authors are in debt to Alan M. da Silva for his relevant contributions. This work was partially supported by Fundação de Amparo à Pesquisa do Estado de São Paulo (FAPESP) and Conselho Nacional de Pesquisas (CNPq).

\appendix

\section{ Barnes-Rivers operators}
\label{ApendiceA}

The $3$-dimensional symmetric Barnes-Rivers operators~\cite{Nieuwenhuizen,Accioly,Rivers} are given by
\bml
\bea
P^1_{ij,kl}&=&2\theta_{((i(k}\omega_{l)j))}\,,\label{23a}\\
P^2_{ij,kl}&=&\theta_{i(k}\theta_{l)j}-\frac{1}{2}\theta_{ij}\theta_{kl}\,,\label{23b}\\
P^0_{ij,kl}&=&\frac{1}{2}\theta_{ij}\theta_{kl}\,,\label{23c}\\
\bar{P}^0_{ij,kl}&=&\omega_{ij}\omega_{kl}\,,\label{23d}\\
\bar{\bar{P}}^0_{ij,kl}&=&\theta_{ij}\omega_{kl}+\omega_{ij}\theta_{kl}\,,\label{23e}
\eea
\eml
where the projection tensors
\be
\label{24}
\theta_{ij}=\delta_{ij}-\frac{k_i k_j}{\vk^2}\quad \mathrm{and}\quad \omega_{ij}=\frac{k_i k_j}{\vk^2}
\ee
have been defined. Such operators obey (using $AB$ in the place of $A^{ij,kl}B_{kl,mn}$ to the contractions) $P^1P^1=P^1$, $P^2P^2=P^2$, $P^0P^0=P^0$, $\bar{P}^0\bar{P}^0=\bar{P}^0$, $\bar{\bar{P}}^0\bar{\bar{P}}^0=(D-1)(P^0+\bar{P}^0)$, $P^0\bar{\bar{P}}^0=\bar{\bar{P}}^0\bar{P}^0=P^{\theta\omega}$, $\bar{P}^0\bar{\bar{P}}^0=\bar{\bar{P}}^0P^0=P^{\omega\theta}$, together with $P^{\theta\omega}_{ij,kl}=\theta_{ij}\omega_{kl}$ and $P^{\omega\theta}_{ij,kl}=\omega_{ij}\theta_{kl}$. Any other contraction is found to be zero. In the present problem, we must also take into account the new object $\mathbb{P}$, whose properties are $k^i\mathbb{P}_{ij,kl}=\theta^{ij}\mathbb{P}_{ij,kl}=0$ while $\theta^i_m\mathbb{P}_{ij,kl}=\mathbb{P}_{mj,kl}$. As a consequence, $\mathbb{P}P=P\mathbb{P}=0$ for $P=P^1,\,P^0,\,\bar{P}^0\,\bar{\bar{P}}^0$ and $\mathbb{P}P^2=P^2\mathbb{P}=\mathbb{P}$. The last important property of $\mathbb{P}$ can be straightforwardly checked and the result is
\be
\label{propQ}
\mathbb{P}_{ij,kl}\mathbb{P}_{kl,mn}=-\frac{1}{8}\vk^{10}P^2_{ij,mn}\,.
\ee
Returning to the Barnes-Rivers operators, they obey the identities
\bml
\bea
\delta_{ij,kl}=(P^1+P^2+P^0+\bar{P}^0)_{ij,kl}\,\label{25a}\\
\delta_{ij}\delta_{kl}=(2P^0+\bar{P}^0+\bar{\bar{P}}^0)_{ij,kl}\,,\label{25b}\\
\frac{4}{\vk^2}\delta_{((i(k}k_{l)}k_{j))}=(2P^1+4\bar{P}^0)_{ij,kl}\,,\label{25c}\\
\frac{1}{\vk^2}(\delta_{ij}k_k k_l+\delta_{kl}k_i k_l)=(\bar{\bar{P}}^0+2\bar{P}^0)_{ij,kl}\,,\label{25d}\\
\frac{1}{\vk^4}(k_i k_j k_k k_l)=\bar{P}^0_{ij,kl}\,.\label{25e}
\eea
\eml
By applying the above identities to the symmetric operator $C$ given in (\ref{2.17}), one can rewrite it as
\be
\label{25-1}
C=x_1P^1+x_2P^2+x_0P^0+\bar{x}_0\bar{P}^0+\bar{\bar{x}}_0\bar{\bar{P}}^0+ix_3\mathbb{P}\,,
\ee
with all the $x$´s real. Let us propose the inverse, if it exists, as
\be
\label{inverse}
\opt^{-1}=y_1P^1+y_2P^2+y_0P^0+\bar{y}\bar{P}^0+\bar{\bar{y}}_0\bar{\bar{P}}^0+y_3\mathbb{P}\,.
\ee
The properties of the $P$´s and the $\mathbb{P}$ together with the result of their products enable one to write the product
\bea
\label{product}
\fl C\opt^{-1}=x_1y_1P^1+\left(x_2y_2-\frac{i}{8}\vk^{10}x_3y_3\right)P^2+(x_0y_0+2\bar{\bar{x}}_0\bar{\bar{y}}_0)P^0
+(\bar{x}_0\bar{y}_0+2\bar{\bar{x}}_0\bar{\bar{y}}_0)\bar{P}^0\nonumber\\
+(x_0\bar{\bar{y}}_0+\bar{\bar{x}}_0\bar{y}_0)P^{\theta\omega}
+(\bar{x}_0\bar{\bar{y}}_0+\bar{\bar{x}}_0y_0)P^{\omega\theta}+(x_2y_3+iy_2x_3)\mathbb{P}=I\nonumber\\
\fl\phantom{C\opt^{-1}}=P^1+P^2+P^0+\bar{P}^0\,,
\eea
where $I$ stands for the identity $\delta^{ij}_{kl}$ obtained in (\ref{25a}). The solution to the above equation then reads
\be
\label{25-2}
\fl \opt^{-1}=\frac{P^1}{x_1}+\frac{x_2P^2}{(x_2)^2-\frac{x_3^2\vk^{10}}{8}}
+\frac{\bar{x}_0P^0+x_0\bar{P}^0-\bar{\bar{x}}_0\bar{\bar{P}}^0}{x_0\bar{x}_0-2\bar{\bar{x}}_0^2}
-i\frac{x_3\mathbb{P}}{(x_2)^2-\frac{x_3^2\vk^{10}}{8}}\,\cdot
\ee

In $4$-dimensions, the Barnes-Rivers operators may be written as
\bml
\bea
\mathcal{P}^1_{\mu\nu,\alpha\beta}&=&2\Theta_{((\mu(\alpha}\Omega_{\beta)\nu))}\,,\label{26a}\\
\mathcal{P}^2_{\mu\nu,\alpha\beta}&=&\Theta_{\mu(\alpha}\Theta_{\beta)\nu}-\frac{1}{3}\Theta_{\mu\nu}\Theta_{\alpha\beta}\,,\label{26b}\\
\mathcal{P}^0_{\mu\nu,\alpha\beta}&=&\frac{1}{3}\Theta_{\mu\nu}\Theta_{\alpha\beta}\,,\label{26c}\\
\bar{\mathcal{P}}^0_{\mu\nu,\alpha\beta}&=&\Omega_{\mu\nu}\Omega_{\alpha\beta}\,,\label{26d}\\
\bar{\bar{\mathcal{P}}}^0_{\mu\nu,\alpha\beta}&=&\Theta_{\mu\nu}\Omega_{\alpha\beta}+\Omega_{\mu\nu}\Theta_{\alpha\beta}\,.\label{26e}
\eea
\eml
Now, the projection operators are defined by
\be
\label{27}
\Theta_{\mu\nu}=\delta_{\mu\nu}-\frac{k_\mu k_\nu}{k^2}\quad\mathrm{and}\quad \Omega_{\mu\nu}=\frac{k_\mu k_\nu}{k^2}\,\cdot
\ee

Eventually, it will be convenient to relate the Barnes-Rivers operators in three and four dimensions. When only the spatial indices are being treated, it is possible to obtain the following identities between the $P$'s and $\mathcal{P}$'s
\bml
\bea
\left(P^1+P^2+P^0+\bar{P}^0\right)_{ij,kl}=\left(\mathcal{P}^1+\mathcal{P}^2+\mathcal{P}^0+\bar{\mathcal{P}}^0\right)_{ij,kl}\,,\label{28a}\\
\left(2P^0+P^0+\bar{\bar{P}}^0\right)_{ij,kl}=\left(3\mathcal{P}^0+\bar{\mathcal{P}}^0+\bar{\bar{\mathcal{P}}}^0\right)_{ij,kl}\,.\label{28b}
\eea
\eml


\end{document}